\documentclass[apsprd,showpacsx]{revtex4}
\usepackage{graphicx}
\usepackage{float}
\usepackage{bm}
\usepackage{diagbox}
\usepackage{amsmath}
\usepackage{array}
\usepackage{color}
\usepackage{eufrak}
\usepackage{mathrsfs}
\usepackage{ulem}

\begin{document}

\title{Meson properties in magnetized quark matter}
\author{Ziyue Wang and Pengfei Zhuang}
\affiliation{Physics Department, Tsinghua University and Collaborative Innovation Center of Quantum Matter, Beijing 100084, China}
\date{\today}
\begin{abstract}
We study neutral and charged meson properties in magnetic field. Taking bosolization method in a two-flavor Nambu--Jona-Lasinio model, we derive effective meson Lagrangian density with minimal coupling to the magnetic field, by employing derivative expansion for both the meson fields and Schwinger phases. We extract from the effective Lagrangian density the meson curvature, pole and screening masses. As the only Goldstone mode, the neutral pion controls the thermodynamics of the system and propagates the lang range quark interaction. The magnetic field breaks down the space symmetry, and the quark interaction region changes from a sphere in vacuum to a ellipsoid in magnetic field.
\end{abstract}
\maketitle

\section{Introduction}
\label{s1}
Strong magnetic field may exist in the core of neutron stars and the beginning of nuclear collisions at high energies. The field strength in these cases is expected to reach $10^{19}$ Gauss, corresponding to $|eB|\sim 10 m_\pi^2$~\cite{Kharzeev:2007jp, Skokov:2009qp, Voronyuk:2011jd}. In such an external magnetic field, the phase structure of a quantum chromodynamics (QCD) system will be significantly changed. For the phase transition from chiral symmetry breaking to its restoration, there are magnetic catalysis effect at mean field level~\cite{Gusynin:1995nb, Klimenko:1991he, Klevansky:1989vi} and inverse magnetic catalysis effect in lattice QCD simulations~\cite{Bali:2011qj, Bali:2012zg, Bali:2013esa} and effective model calculations~\cite{Fukushima:2012kc,Chao:2013qpa,Bruckmann:2013oba,Andersen:2014oaa,Mao:2016fha}. Considering that pion mesons are the Goldstone modes corresponding to chiral symmetry breaking and dominate the QCD thermodynamics at low temperature, their properties~\cite{Fayazbakhsh:2013cha,Avancini:2015ady,Avancini:2016fgq,Mao:2017wmq,Fayazbakhsh:2012vr,Kamikado:2013pya,Andersen:2012dz,Andersen:2012zc,Colucci:2013zoa,
Hattori:2015aki,Mukherjee:2017dls} in an external magnetic field are extensively investigated.

In quark models mesons are considered as quantum fluctuations. They are usually constructed through non-perturbative methods like random phase approximation~\cite{Avancini:2015ady,Mao:2017wmq} and bosolization~\cite{Eguchi:1976iz,Klevansky:1992qe,Imai:2012hr}. One problem related to constructing charged mesons in magnetic field in these models is the treatment of Schwinger phase. The problem comes from the lack of translation invariance of the Schwinger phase in quark propagator. For neutral meson polarization functions the two Schwinger phases for the quark-antiquark pair cancel to each other, but for charged mesons one cannot simply perform a Fourier transformation to obtain polarization functions in momentum space. A direct way to deal with the problem is to work in coordinate space all the way, but the calculation is expected to be of great complexity. In some of the calculations, people simply neglect the Schwinger phase and take only the invariant part~\cite{Liu:2016vuw,Zhang:2016qrl}.

In this paper, we study meson properties in magnetized quark matter in the frame of a two-flavor Nambu--Jona-Lasinio model. After a simple introduction of the mean field calculation in the model in Section \ref{s2}, we focus on deriving effective meson Lagrangian density with minimal coupling to the magnetic field, by taking derivative expansion for both the meson fields and Schwinger phases in Section \ref{s3}. We extract from the meson Lagrangian density the meson masses and wave function renormalization constants in this section. We show numerical calculations for the meson curvature, pole and screening masses at finite temperature and magnetic field in Section \ref{s4}. We summarize in Section \ref{s5}.

\section{Mean Field Approximation}
\label{s2}

Nambu--Jona-Lasinio (NJL) models at quark level describe well the chiral symmetry breaking in vacuum and its restoration at finite temperature and baryon density~\cite{Nambu:1961tp,Klevansky:1992qe,Hatsuda:1994pi}. In the models, mesons are treated as collective excitations of quarks, and the magnetic effect on mesons comes from the magnetized quarks. In the presence of an external electromagnetic field the two-flavor NJL model is defined as
\begin{equation}
\mathcal{L}=\bar{\psi}(i\gamma^\mu D_\mu-m_0)\psi+G\Big[(\bar{\psi}\psi)^2+(\bar{\psi}i\gamma_5{\bf \tau}\psi)^2\Big],
\end{equation}
where $D_\mu=\partial_\mu+iQ A_\mu$ is the covariant derivative with the electromagnetic potential $A_\mu=(0,0,Bx,0)$ which gives a constant magnetic field along the $z$ axis ${\bf B}=B{\bf e}_z$, $Q=diag(Q_u=2e/3,Q_d=-e/3)$ the diagonal quark charge matrix in flavor space, $m_0$ the current quark mass characterizing the explicit chiral symmetry breaking, ${\bf \tau}$ the Pauli matrices in isospin space, and $G$ the quark coupling constant in scalar and pseudoscalar channels.

Introducing scalar and pseudoscalar meson fields $M(x) = (\sigma(x),{\bf \pi}(x)) = (-2G\bar{\psi}(x)\psi(x), -2G\bar{\psi}(x)i\gamma^5{\bf \tau}\psi(x))$, and integrating out the fermion fields, the effective Lagrangian density of the model is expressed as,
\begin{equation}
\mathcal{L} = -\sum_M{\left(g_M M\right)^2\over 4G}-i\text{Tr} \ln \left(i\gamma^\mu D_\mu-m_0-K\right),
\label{action}
\end{equation}
where $K=\sum_M \Gamma_M M$ shows the interaction between mesons and quarks, $\Gamma_M=(\Gamma_\sigma,\Gamma_\pi)=(g_\sigma,ig_\pi\gamma_5{\bf \tau})$ are the interaction vertexes with meson coupling constants $g_\sigma$ and $g_\pi$, and $\text{Tr}$ means the trace in inner spaces like color, flavor and spin.

Introducing the ensemble average $\langle\sigma\rangle$ which is the order parameter of the chiral phase transition, making a shift $\sigma(x)\rightarrow \langle\sigma\rangle +\sigma(x)$, and keeping only the condensate $\langle\sigma\rangle$ and dropping all the meson fluctuations, one obtains the mean field Lagrangian density
\begin{equation}
\mathcal{L}_{\text{MF}}=-i\text{Tr} \ln S^{-1}-{(m-m_0)^2\over 4G}
\end{equation}
and the corresponding thermodynamical potential
\begin{equation}
\Omega_{\text{MF}} = {(m-m_0)^2\over 4G}+i\frac{T}{V}\sum_x\text{Tr} \ln S^{-1}
\end{equation}
with the definition $\sum_x=\int d^4 x$ ($\sum_p=\int d^4p/(2\pi)^4$), where $m=m_0+g_\sigma\langle\sigma\rangle$ is the dynamic quark mass, and $S=(i\gamma^\mu D_\mu-m)^{-1}=diag(S_u,\ S_d)$ is the quark propagator in mean field approximation.

The physical value of the chiral condensate $\langle\sigma\rangle$ or the quark mass $m$ is determined by the minimum of the thermodynamical potential $\partial\Omega_\text{MF}/\partial\langle\sigma\rangle = 0$, namely the gap equation
\begin{equation}
\label{gap}
m = m_0+2iG \sum_x\text{Tr}S = m_0+2iG N_c\sum_{q,p}\text{Tr} \widetilde{S}_q(p),
\end{equation}
where $\widetilde{S}_q(p)$ is the Fourier transformation of the translation invariant part $\widetilde{S}_q(x-y)$ of the quark propagator $S_q(x,y)=e^{i\Phi_q(x,y)}\widetilde{S}_q(x-y)$ defined in Schwinger formalism~\cite{Gusynin:1995nb},
\begin{eqnarray}
\widetilde{S}_q &=& i e^{-p_\perp^2/|Q_qB|}\sum_{n=0}^{\infty}(-1)^n\frac{D_{n}(p)}{G_{n}(p)},\\
G_{n} &=& p_0^2-2n|Q_qB|-p_3^2-m^2,\nonumber\\
D_{n} &=& \left(p_0\gamma_0-p_3\gamma_3+m\right)\left[(1-is_q\gamma_1\gamma_2)L_n(p_q)-(1+is_q\gamma_1\gamma_2)L_{n-1}(p_q)\right]+4(p_1\gamma_1+p_2\gamma_2)L^1_{n-1}(p_q)\nonumber
\end{eqnarray}
with $s_q=sgn(Q_qB)$, $p_q=2p_\perp^2/|Q_qB|$ and Laguerre polynomials $L_n(z)$. Note that the Schwinger phase $\Phi_q(x,y)$ in the quark propagator is not translation invariant.

Using the orthogonal relationship for the Laguerre polynomials and performing Matsubara frequency summation at finite temperature, the gap equation can be explicitly written as an integral over the momentum along the field direction and a summation over the Landau energy level,
\begin{eqnarray}
\label{i2}
m &=& m_0+4m  G I_2,\\
I_2 &=& N_c\sum_{q,n,p_3}\frac{|Q_qB|}{2\pi}\alpha_n{1-2n_f(E_{q,n})\over 2E_{q,n}}\nonumber
\end{eqnarray}
with the spin degeneracy factor $\alpha_n=2-\delta_{n0}$, the Fermi-Dirac distribution $n_f(x)=1/(1+e^{x/T})$ and the quark energy $E_{q,n}=\sqrt{2n|Q_qB|+p_3^2+m^2}$.

\section{Meson Properties from Bosonization}
\label{s3}
We now go beyond the mean field approximation. Including the $\sigma$ and $\pi$ fluctuations, the effective Lagrangian density (\ref{action}) can be formally written as
\begin{equation}
\label{seff}
\mathcal{L}=\mathcal{L}_{\text{MF}}+\mathcal{L}_{\text{FD}}-{2(m-m_0)\sigma+\sum_M (g_M M)^2\over 4G}.
\end{equation}
The fermionic determinant part $\mathcal{L}_{\text{FD}}$ contains highly non-local interactions. It is desirable to localize it in a systematic fashion, and one such procedure is the derivative expansion of the fields~\cite{Gusynin:1995nb,Fayazbakhsh:2012vr,Eguchi:1976iz,Klevansky:1992qe,Imai:2012hr,Zhang:2016qrl}. To this end, we first make Taylor expansion,
\begin{eqnarray}
\mathcal{L}_{\text{FD}} &=& -i\text{Tr} \ln (1-SK)=\sum_{n=1}^\infty {\cal L}^{(n)},\nonumber\\
{\cal L}^{(n)} &=& \frac{i}{n}\text{Tr}(SK)^n.
\end{eqnarray}

We first work out the first term ${\cal L}^{(1)}=i\text{Tr}(SK)$, which is the linear term in the scalar meson field,
\begin{equation}
{\cal L}^{(1)} = i \text{Tr}(S_u+S_d)g_\sigma\sigma=iN_c\sum_{q,p}\text{Tr}\widetilde{S}_q(p)g_\sigma\sigma.
\end{equation}
Combining with the other linear term in ${\cal L}$, the disappearance of the whole linear term from the effective Lagrangian recovers the gap equation (\ref{gap}) for the quark mass.

The meson kinetic energies and masses are extracted from the second term with $n=2$ in ${\cal L}_\text{FD}$, ${\cal L}^{(2)}=i/2\text{Tr}(SK)^2$. The trace in flavor space gives
\begin{eqnarray}
\label{u2}
\text{Tr}_q(SK)^2 &=& g_\sigma^2\left(S_u \sigma S_u \sigma+S_d \sigma S_d \sigma\right)+g_\pi^2\left(S_u i\gamma_5\pi_0 S_u i\gamma_5 \pi_0+S_d i\gamma_5 \pi_0 S_d i\gamma_5 \pi_0\right)\nonumber\\
&& +2g_\pi^2\left( S_u i\gamma_5 \pi_-S_d i\gamma_5 \pi_+ +S_d i\gamma_5 \pi_+S_u i\gamma_5 \pi_-\right).
\end{eqnarray}

For the $\sigma$ part we have
\begin{eqnarray}
{\cal L}_\sigma^{(2)} &=& {ig_\sigma^2 N_c\over 2} \sum_{q,y}e^{i\left(\Phi_q(x,y)+\Phi_q(y,x)\right)}\text{Tr}\widetilde{S}_q(x-y)\widetilde{S}_q(y-x)\sigma(y)\sigma(x)\nonumber\\
&=& {ig_\sigma^2 N_c\over 2} \sum_{q,y} \text{Tr}\widetilde{S}_q(x-y)\widetilde{S}_q(y-x)\sigma(y)\sigma(x),
\end{eqnarray}
where the two Shwinger phases cancel to each other, $\Phi_q(x,y)+\Phi_q(y,x)=0$, and only the translation invariant part of the quark propagator contributes to the action. We then take a local expansion of the $\sigma$ field at $x$,
\begin{equation}
\sigma(y) = \sigma(x)+(y-x)^\mu\partial_\mu \sigma(x)+\frac{1}{2}(y-x)^\mu(y-x)^\nu \partial_\mu\partial_\nu \sigma(x)+\cdots.
\end{equation}
By substituting this expansion into ${\cal L}_\sigma^{(2)}$, the first term which is proportional to $\sigma^2(x)$ contributes to the meson curvature masses,
\begin{equation}
\label{mass1}
\overline m_\sigma^2 = -ig_\sigma^2 N_c\sum_{q,p}\text{Tr}\widetilde{S}_q(p)\widetilde{S}_q(p) = -2g_\sigma^2\left(I_2-2m^2I_0\right),
\end{equation}
where the integral $I_2$ is shown in (\ref{i2}) and $I_0$ is defined by
\begin{equation}
I_0 = N_c\sum_{q,n,p_3}\frac{|Q_qB|}{2\pi}\alpha_n{1-2n_f(E_{q,n})+2E_{q,n}n_f'(E_{q,n})\over 4E_{q,n}^3}
\end{equation}
with the notation $n_f'(x)=dn_f(x)/dx$. Note that, $\overline m_\sigma$ is only a part of the curvature mass, the other quadratic term in ${\cal L}$ contributes to the curvature mass too.

The second term in ${\cal L}^{(2)}_\sigma$ which is linear in $(y-x)_\mu$ is the surface term and can be shown to vanish, and the third term $\sim \sigma(x)\partial_\mu\partial_\nu \sigma(x)$ gives the kinetic energy. Performing a Fourier transformation for the translation invariant part of the quark propagator and doing partial integration, the third term becomes
\begin{equation}
{ig_\sigma^2 N_c\over 4} \sum_{q,y}\text{Tr}\widetilde{S}_q(x-y)\widetilde{S}_q(y-x)\sigma(x)(y-x)^\mu(y-x)^\nu \partial_\mu\partial_\nu \sigma(x)=\frac{1}{2}\mathcal{F}_\sigma^{\mu\nu}\partial_\mu \sigma(x)\partial_\nu \sigma(x),
\end{equation}
with the coefficients
\begin{equation}
\mathcal{F}_\sigma^{\mu\nu}=-{ig_\sigma^2 N_c\over 2} \sum_{q,p,k,z} \text{Tr} e^{i(p-k)z}z^\mu z^\nu\widetilde{S}_q(p)\widetilde{S}_q(k),
\end{equation}
which can be considered as wave function renormalization constants for the $\sigma$ field. Using the relations $z_\mu e^{i(k-p)z}=i\partial_\mu^p e^{i(k-p)z}$ and $z_\mu z_\nu e^{i(k-p)z}=-\partial_\nu^p\partial_\mu^p e^{i(k-p)z}$, we have
\begin{equation}
\mathcal{F}_\sigma^{\mu\nu} = {ig_\sigma^2 N_c\over 2} \sum_{q,p} \text{Tr}\left[\widetilde{S}_q(p)\frac{\partial^2}{\partial p_\mu\partial p_\nu}\widetilde{S}_q(p)\right].
\end{equation}
The coefficients $\mathcal{F}_\sigma^{\mu\nu}$ with $\mu\neq\nu$ can be shown to vanish, and the nontrivial elements are only the diagonal ones $\mathcal{F}_\sigma^{00}$, $\mathcal{F}_\sigma^{33}$, $\mathcal{F}_\sigma^{11}$ and $\mathcal{F}_\sigma^{22}$. While the system is no longer isotropic in a background magnetic field, the coefficients in the directions perpendicular to the field should be the same, $\mathcal{F}_\sigma^{11} = \mathcal{F}_\sigma^{22}$. After taking trace in Dirac space, integrating over $p_1$ and $p_2$, and making a Wick rotation, we have the wave function renormalization constants
\begin{eqnarray}
\label{fs}
\mathcal{F}_{\sigma}^{00} &=& \frac{g_\sigma^2 N_c}{\pi}\sum_{q,n,p_0,p_3}|Q_qB|\alpha_n{(p_0^2+E_{q,n}^2)^2-2(p_0^2+m^2)(p_0^2+E_{q,n}^2)+8m^2p_0^2\over (p_0^2+E_{q,n}^2)^4},\nonumber\\
\mathcal{F}_{\sigma}^{33} &=& \frac{g_\sigma^2 N_c}{\pi}\sum_{q,n,p_0,p_3}|Q_qB|\alpha_n{(p_0^2+E_{q,n}^2)^2-2(p_3^2+m^2)(p_0^2+E_{q,n}^2)+8m^2p_3^2\over (p_0^2+E_{q,n}^2)^4},\nonumber\\
\mathcal{F}_{\sigma}^{11} &=& -\frac{g_\sigma^2 N_c}{2\pi} \sum_{q,i,j,p_0,p_3}(-1)^{i+j}\frac{(p_0^2+p_3^2-m^2)J_1(i,j)+4|Q_qB|J_2(i,j)}{(p_0^2+E_{q,i}^2)(p_0^2+E_{q,j}^2)},
\end{eqnarray}
where $J_1$ and $J_2$ are integrals defined as
\begin{eqnarray}
J_1(i,j) &=& \sum_{k=0}^1\int_0^\infty du e^{-u} L_{i-k}(u)\left[(u-2)L_{j-k}(u)+4(u-1)L_{j-k-1}^1(u)~+~4uL_{j-k-2}^2(u)\right],\nonumber\\
J_2(i,j) &=& \int_0^\infty du e^{-u}uL_{i-1}^1(u)\left[(u-4)L_{j-1}^1(u)+4(u-2)L_{j-2}^2(u)+4uL_{j-3}^3(u)\right].
\end{eqnarray}
From the orthogonal normalization of the Laguerre polynomials, $J_1$ and $J_2$ can be explicitly worked out,
\begin{eqnarray}
J_1(i,j) &=& -(2i-1)\delta_{j,i-1}-(2i+1)\delta_{j,i+1}-4i\delta_{j,i},\quad \text{for}\ i\geq 1,\nonumber\\
J_1(0,0) &=& J_1(0,1)=-1,\quad \text{for}\ i=0,\nonumber\\
J_2(i,j) &=& -i(i-1)\delta_{j,i-1}-i(i+1)\delta_{j,i+1}-2i^2\delta_{j,i},\quad \text{for}\ i\geq 1.
\end{eqnarray}
Note that, the integral over $p_0$ in (\ref{fs}) means the Matsubara frequency summation in the imaginary time formalism of finite temperature field theory.

For the neutral pion field $\pi_0$, we can straightforwardly treat its contribution to ${\cal L}^{(2)}$,
\begin{equation}
{\cal L}_{\pi_0}^{(2)} = {ig_\pi^2 N_c\over 2} \sum_{q,y} \text{Tr}\widetilde{S}_q(x-y)i\gamma_5\widetilde{S}_q(y-x)i\gamma_5\pi_0(y)\pi_0(x),\nonumber
\end{equation}
by taking similar calculation as for $\sigma$ field, and extract its contribution to the curvature mass and the wave function renormalization constants,
\begin{eqnarray}
\label{mass2}
\overline m_{\pi_0}^2 &=& -2g_\pi^2 I_2,\nonumber\\
\mathcal{F}_{\pi_0}^{00} &=& \frac{g_\pi^2 N_c}{\pi}\sum_{qn,p_0,p_3}|Q_qB|\alpha_n{E_{q,n}^2-p_0^2\over (p_0^2+E_{q,n}^2)^3},\nonumber\\
\mathcal{F}_{\pi_0}^{33} &=& \frac{g_\pi^2 N_c}{\pi}\sum_{q,n,p_0,p_3}|Q_qB|\alpha_n{p_0^2+E_{q,n}^2-2p_3^2\over (p_0^2+E_{q,n}^2)^3},\nonumber\\
\mathcal{F}_{\pi_0}^{11} &=& -\frac{g_\pi^2 N_c}{2\pi} \sum_{q,i,j,p_0,p_3} (-1)^{i+j}\frac{(p_0^2+p_3^2+m^2)J_1(i,j)+4|Q_qB|J_2(i,j)}{\big(p_0^2+E_{q,i}^2\big)\big(p_0^2+E_{q,j}^2\big)}.
\end{eqnarray}

We now calculate the charged meson part of the effective Lagrangian ${\cal L}^{(2)}$. Different from neutral mesons, the two Schwinger phases appeared in the quark loop do not cancel to each other and have contribution to the Lagrangian. Taking the full quark propagator, the $\pi_-$ part is written as
\begin{equation}
{\cal L}_{\pi_-}^{(2)} =ig_\pi^2 N_c\sum_y e^{i\Phi_{\pi_-}(y,x)}\text{Tr}\widetilde{S}_u(x-y) i\gamma^5 \widetilde{S}_d(y-x) i\gamma^5\pi_-(y)\pi_+(x)
\end{equation}
with the Schwinger phase
\begin{equation}
\Phi_{\pi_-}(y,x) = \Phi_u(x,y)+\Phi_d(y,x) = Q_u\int_y^xA^\mu(x')d x'_\mu+Q_d\int_x^yA^\mu(x')d x'_\mu = -e\int_x^yA^\mu(x')d x'_\mu.
\end{equation}

By taking local expansion for both the charged field and the phase,
\begin{eqnarray}
\pi_-(y) &=& \pi_-(x)+(y-x)^\mu\partial_\mu \pi_-(x) +\frac{1}{2}(y-x)^\mu (y-x)^\nu \partial_\mu\partial_\nu \pi_-(x)+\cdots,\nonumber\\
e^{i\Phi_{\pi_-}(y,x)} &=& 1-ie A^\mu(x) (y-x)_\mu- i e \frac{1}{2}\frac{\partial A^\mu(x)}{\partial x^\nu} (y-x)_\mu  (y-x)_\nu+\frac{(-i e)^2}{2}\left(A^\mu(x) (y-x)_\mu\right)^2+\cdots,
\end{eqnarray}
${\cal L}_{\pi_-}^{(2)}$ up to the quadratic term is expressed as
\begin{eqnarray}
{\cal L}_{\pi_-}^{(2)} &=& ig_\pi^2 N_c\sum_y\text{Tr}\widetilde{S}_u(x-y)i\gamma^5\widetilde{S}_d(y-x)i\gamma^5\nonumber\\
&&\times\left[\pi_-(x)+(y-x)^\mu D^-_\mu\pi_-(x)+\frac{1}{2}(y-x)^\mu (y-x)^\nu D^-_\mu  D^-_\nu\pi_-(x)\right]\pi_+(x)
\end{eqnarray}
with the covariant derivative $D^-_\mu = \partial_\mu - ie A_\mu(x)$. Taking almost the same calculation for the positively charged pion field, we have
\begin{eqnarray}
{\cal L}_{\pi_+}^{(2)} &=& ig_\pi^2 N_c\sum_y\text{Tr} \widetilde{S}_d(x-y)i\gamma^5\widetilde{S}_d(y-x)i\gamma^5\nonumber\\
&&\times\left[\pi_+(x)+(y-x)^\mu D^+_\mu\pi_+(x)+\frac{1}{2}(y-x)^\mu (y-x)^\nu D^+_\mu  D^+_\nu\pi_+(x)\right]\pi_-(x)
\end{eqnarray}
with the covariant derivative $D^+_\mu = \partial_\mu + ie A_\mu(x)$. The two terms linear in $(y-x)^\mu$ in ${\cal L}_{\pi_-}^{(2)}$ and ${\cal L}_{\pi_+}^{(2)}$ can be summed up to give a surface term which vanishes in the sense of coordinate integration. We finally write the $\pi_\pm$ contribution to the effective Lagrangian in terms of the masses $\overline m_{\pi_\pm}$ and the wave function renormalization constants ${\cal F}_{\pi_\pm}^{\mu\mu}$ under the current gauge $\partial^\mu A_\mu=0$,
\begin{equation}
{\cal L}_{\pi_+}^{(2)}+{\cal L}_{\pi_-}^{(2)}=\frac{1}{2}\left(\mathcal{F}^{\mu\mu}_{\pi_+}\big|D_\mu\pi_+\big|^2+\mathcal{F}^{\mu\mu}_{\pi_-}\big|D_\mu\pi_-\big|^2\right)- \frac{1}{2}\left(\overline m_{\pi_+}^2\big|\pi_+\big|^2+\overline m_{\pi_-}^2\big|\pi_-\big|^2\right),
\end{equation}
with the coefficients
\begin{eqnarray}
\overline m_{\pi_-}^2 &=& -\frac{2g_\pi^2N_c|Q_uB|}{\pi}\sum_{i,j,p_0,p_3}(-1)^{i+j}{Y_{12}\over (p_0^2+E_{d,j}^2)(p_0^2+E_{u,i}^2)},\nonumber\\
\mathcal{F}^{00}_{\pi_-} &=& \frac{2g_\pi^2N_c|Q_uB|}{\pi}\sum_{i,j,p_0,p_3}(-1)^{i+j}{Y_{12}\left(E_{u,i}^2-3p_0^2\right)+2p_0^2Y_1(p_0^2+E_{u,i}^2)\over (p_0^2+E_{u,i}^2)^3(p_0^2+E_{d,j}^2)},\nonumber\\
\mathcal{F}^{33}_{\pi_-} &=& \frac{2g_\pi^2N_c|Q_uB|}{\pi}\sum_{i,j,p_0,p_3}(-1)^{i+j}{Y_{12}\left(p_0^2+E_{u,i}^2-4p_3^2\right)+2p_3^2Y_1(p_0^2+E_{u,i}^2)\over (p_0^2+E_{u,i}^2)^3(p_0^2+E_{d,j}^2)},\nonumber\\
\mathcal{F}^{11}_{\pi_-} &=& \frac{g_\pi^2N_c}{\pi}\sum_{i,j,p_0,p_3}(-1)^{i+j}\frac{H_{12}}{(p_0^2+E_{d,j}^2)(p_0^2+E_{u,i}^2)},
\end{eqnarray}
where the integrals $Y_1$, $Y_2$, $H_1$, $H_2$, $Y_{12}$ and $H_{12}$ are defined as
\begin{eqnarray}
Y_1(i,j) &=& \int_0^\infty du e^{-3u/2}\left[L_i(u)L_{j-1}(2u)+L_{i-1}(u)L_{j}(2u)\right],\nonumber\\
Y_2(i,j) &=& \int_0^\infty du e^{-3u/2} u  L_{i-1}^1(u)L_{j-1}^1(2u),\nonumber\\
H_1(i,j) &=& \sum_{k=0}^1\int_0^\infty du e^{-3u/2} L_{j-k}(2u)\left[(u-2)L_{i-k}(u)+4(u-1)L_{i-k-1}^1(u)+4uL_{i-k-2}^2(u)\right],\nonumber\\
H_2(i,j) &=& \int_0^\infty du~e^{-3u/2}uL_{j-1}^1(2u)\left[(u-4)L_{i-1}^1(u)+4(u-2)L_{i-2}^2(u)+4uL_{i-3}^3(u)\right],\nonumber\\
Y_{12} &=& (p_0^2+p_3^2+m^2)Y_1-4|Q_uB|Y_2,\nonumber\\
H_{12} &=& (p_0^2+p_3^2+m^2)H_1-4|Q_uB|H_2.
\end{eqnarray}
Due to the symmetry between the quark loops for $\pi_+$ and $\pi_-$, it is easy to show $\overline m_{\pi_+}=\overline m_{\pi_-}$ and ${\cal F}_{\pi_+}^{\mu\mu} = {\cal F}_{\pi_-}^{\mu\mu}$. Note that, including self-consistently the Schwinger phases $\Phi_{\pi_\pm}$ in the calculation leads to the covariant derivatives $D^\pm_\mu$ acting on the charged pion fields, which guarantee the minimal coupling between the meson fields and the gauge field in the frame of bosonization.

Putting together all the contributions from neutral and charged mesons ${\cal L}^{(2)}_\sigma$, ${\cal L}^{(2)}_{\pi_0}$, ${\cal L}^{(2)}_{\pi_+}$ and ${\cal L}^{(2)}_{\pi_-}$ calculated above, we obtain the effective Lagrangian of the quark-meson plasma at quasiparticle level,
\begin{equation}
\label{seff2}
\mathcal{L} = \mathcal{L}_{\text{MF}}+\sum_M\left[\frac{1}{2}\mathcal{F}_M^{\mu\mu}|{\cal D}^M_\mu M(x)|^2-\frac{1}{2}\left(\frac{g_M^2}{2G}+\overline m_M^2\right)|M(x)|^2\right]
\end{equation}
with ${\cal D}^M_\mu=\partial_\mu$ for neutral mesons and $D_\mu^\pm=\partial_\mu\pm ieA_\mu$ for charged mesons, where we have neglected the linear term in $\sigma$ and all the surface terms. The effective action (\ref{seff2}) defines the meson curvature masses
\begin{equation}
\label{mass3}
m_M^2={g_M^2\over 2G}+\overline m_M^2.
\end{equation}
For the neutral pion, we have from (\ref{mass2}) $m_{\pi_0}^2=g_0^2/(2G)(1-4GI_2)$. The comparison with the quark gap equation (\ref{i2}) leads to $m_{\pi_0}=0$ in the chiral breaking phase in chiral limit with $m_0=0$. This guarantees the Goldstone mode in finite temperature and magnetic field. Note that, the original $SU_L(2)\times SU_R(2)$ chiral symmetry is broken down to $U_L(1)\times U_R(1)$ by the magnetic field ${\bf B}$ and the
number of Goldstone modes is reduced from 3 to 1.

From the on-shell conditions for noninteracting mesons in an external magnetic field,
\begin{eqnarray}
&& {\mathcal{F}^{00}_M}p_0^2 -\mathcal{F}^{11}_M p_1^2-\mathcal{F}^{22}_M p_2^2-\mathcal{F}^{33}_M p_3^2 = m_M^2,\quad \text{for}\ M=\sigma,\pi_0\nonumber\\
&& {\mathcal{F}^{00}_M}p_0^2-\mathcal{F}^{11}_M(2n+1)|eB|-\mathcal{F}^{33}_M p_3^2 = m_M^2\quad \text{for}\ M=\pi_+,\pi_-,
\label{dispersion}
\end{eqnarray}
we obtain the meson pole masses defined as $p_0=m_M^{(0)}$ at zero momentum $p_1=p_2=p_3=0$ for neutral mesons and at lowest Landau level and zero momentum $n=p_3=0$ for charged mesons,
\begin{eqnarray}
\label{pole}
m^{(0)}_M &=& m_M/\sqrt{\mathcal{F}^{00}_M}\quad \text{for}\ M=\sigma,\pi_0\nonumber\\
m^{(0)}_M &=& \sqrt{\left(m_M^2+\mathcal{F}^{11}_M|eB|\right)/{\cal F}^{00}_M}\quad \text{for}\ M=\pi_+,\pi_-.
\end{eqnarray}

We can also define the screening masses at $p_j=i m_M^{(j)}$ in the direction ${\bf e}_j$. For instance, we have for neutral mesons,
\begin{equation}
m_M^{(j)}=m_M/\sqrt{\mathcal{F}^{jj}_M}\quad \text{for}\ M=\sigma,\pi_0.
\end{equation}
Since the two directions perpendicular to the magnetic field are symmetric, we have only two independent screening masses, the parallel and perpendicular screening masses $m_M^{(z)}$ and $m_M^{(x)}$. By considering the interaction potential between two quarks which is just a Fourier transform of the meson propagator, the screening mass $m_M^{(j)}$ controls the interaction range, its inverse defines the screening radius,
\begin{equation}
r^{(j)}_M=1/m^{(j)}_M.
\end{equation}
It is clear that, heavy mesons like the Higgs mode $\sigma$ propagate short range interaction and light mesons like the Goldstone mode $\pi_0$ propagate the long range interaction.

\section{Numerical Results}
\label{s4}
Before we do numerical calculation in finite temperature and magnetic field, we first fix the model parameters by fitting the meson properties in vacuum. Since temperature and magnetic field disappear in vacuum, there is no difference among the three pions ($m_\pi = m_{\pi_0}=m_{\pi_\pm}$), and the symmetry in space-time is recovered ($\mathcal{F}_M=\mathcal{F}_M^{00}=\mathcal{F}_M^{33}=\mathcal{F}_M^{11}=\mathcal{F}_M^{22}$). Taking into account the gap equation for the quark mass $m$ (\ref{i2}) and the pion decay constant $f_\pi$ in the model \cite{Klevansky:1992qe} $f_\pi^2=m^2I_0$, the meson curvature masses (\ref{mass3}) can be expressed in terms of $m$ and $f_\pi$,
\begin{eqnarray}
m_\pi^2 &=& {g_\pi^2\over 2G}{m_0\over m},\nonumber\\
m_\sigma^2 &=& {g_\sigma^2\over 2G}\left({m_0\over m}+8Gf_\pi^2\right).
\end{eqnarray}

On the other hand, the meson wave function renormalization constants to the leading order are greatly simplified in vacuum,
\begin{eqnarray}
\mathcal{F}_\sigma &\simeq& g_\sigma^2 I_0 = g_\sigma^2{f_\pi^2\over m^2},\nonumber\\
\mathcal{F}_\pi &\simeq& g_\pi^2 I_0 = g_\pi^2{f_\pi^2\over m^2}.
\end{eqnarray}
With the Goldberger-Treiman relation \cite{Klevansky:1992qe} $f_\pi^2g_\pi^2=m^2$, we have
\begin{equation}
\mathcal {F}_\sigma = \mathcal {F}_\pi =1
\end{equation}
in vacuum. This means that, the wave function renormalization is not important in vacuum and we can safely neglect it. In this case, we have the meson pole masses
\begin{equation}
m^{(0)}_M \simeq m_M^{(z)} \simeq m_M^{(x)} \simeq m_M.
\end{equation}

As a non-renormalizable theory, the results of the NJL model are regularization scheme dependent. We adopt a Pauli-Villars regularization~\cite{Klevansky:1992qe, Mao:2016fha} in our numerical calculation in order to avoid unphysical problems caused by simple hard and soft cutoff schemes~\cite{Fayazbakhsh:2013cha}. Under the Pauli-Villars regularization, one replaces any integrated function $G(m)$ by $\sum_i c_i G(m_i)$ with regularized quark masses $m_i=\sqrt{m^2+a_i\Lambda^2}$ for $i=0,1,\cdots,N-1$. The coefficients $a_i$ and $c_i$ are determined by constraints $a_0 =0$, $c_0=1$ and $\sum_{i=0}^{N-1} c_im_i^{2j}=0$ for $j=0,1,\cdots, N-1$. In our treatment we take a $N=8$ Pauli-Villars regularization with $a_i=\{0,1,2,3,4,5,6,7\}$ and $c_i=\{1, -7, 21, -35, 35, -21, 7, -1\}$. The parameters $\Lambda=1.505\ \text{GeV}$, $G= 3.44\ (\text{GeV})^{-2}$ and $m_0=5.3\ \text{MeV}$ are obtained by fitting $f_\pi=93\ \text{MeV}$, $\langle\bar\psi\psi\rangle=(-246\ \text{MeV})^3$ and $m_\pi=135\ \text{MeV}$. Note that, in the Pauli-Villars regularization scheme the summation over Landau energy level converges slower than that in hard and soft cutoff schemes. In our calculation the summation is performed up to $n=2000$ for $eB=2m_\pi^2$, to guarantee the convergence of the calculation.

\begin{figure}[H]
\centering
\includegraphics[height=0.50\textwidth]{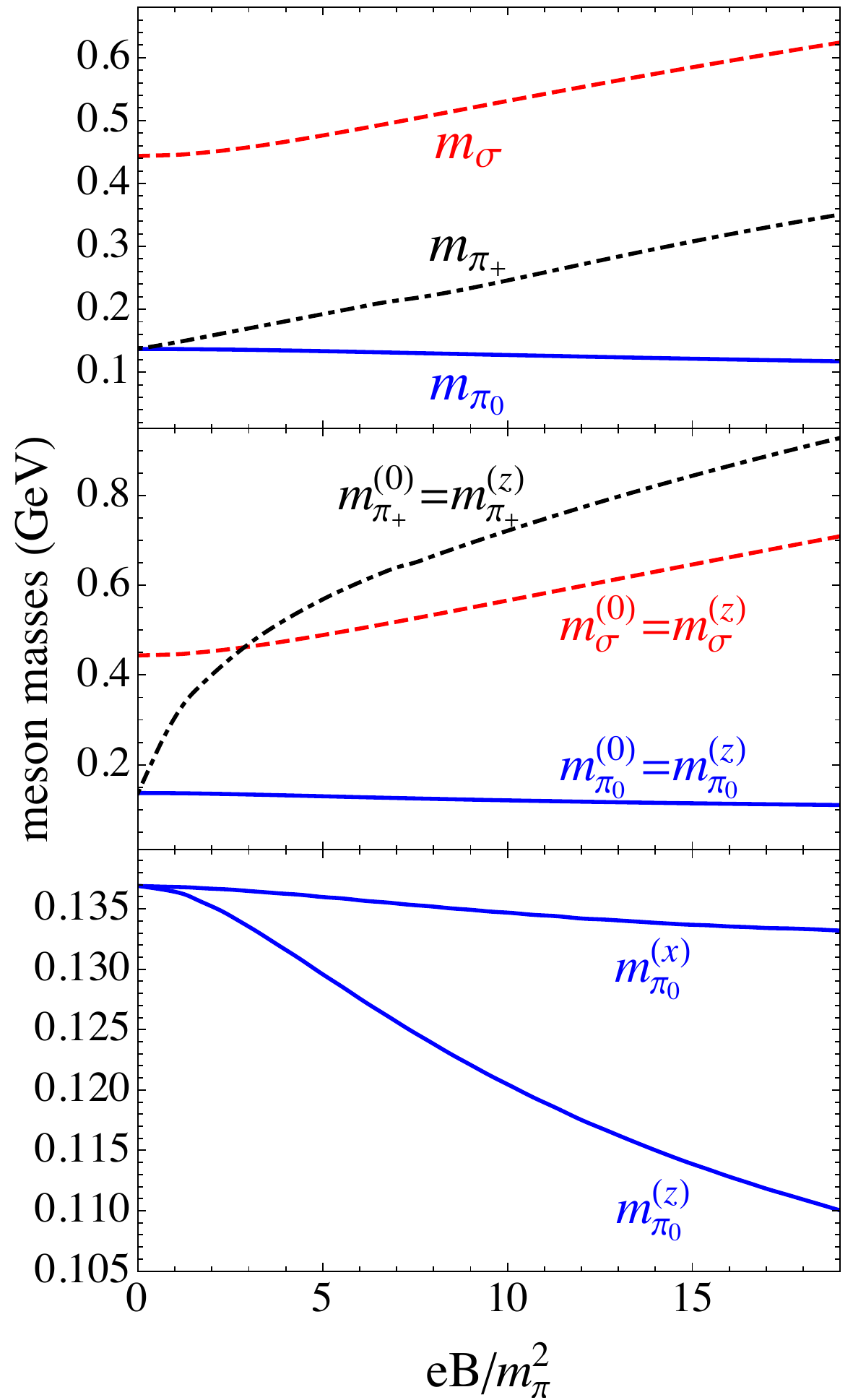}
\caption{The meson curvature masses (upper panel) and pole and screening masses (middle and lower panels) at zero temperature but finite magnetic field. $m_\pi$ is the pion mass in vacuum. }
\label{fig1}
\end{figure}
We first show in Fig.\ref{fig1} the meson curvature, pole and screening masses as functions of magnetic field at zero temperature. Considering the Lorentz invariance between the time direction and magnetic field direction at zero temperature, there is no difference between the wave function renormalizations in these two directions. Therefore, the pole mass $m_M^{(0)}$ is exactly the same as the parallel screening mass $m_M^{(z)}$. In the beginning at zero magnetic field, the difference between the parallel and perpendicular directions disappears too, and the wave function renormalization can be approximately neglected. In this case, the only mass difference is between the isospin singlet and triplet, the degenerated pion mass starts at 135 MeV and sigma mass at 440 MeV. The nonzero magnetic field brings a mass splitting between neutral and charged pions: The curvature masses $m_{\pi_+}=m_{\pi_-}$ goes up but $m_{\pi_0}$ goes down with increasing $eB$. For neutral mesons, the correction from wave function renormalizations in the time and $z$ directions is still weak when the magnetic field is not very strong, which leads to almost the same curvature, pole and screening masses $m_\sigma^{(0)}=m_\sigma^{(z)}\simeq m_\sigma$ and $m_{\pi_0}^{(0)}=m_{\pi_0}^{(z)}\simeq m_{\pi_0}$. For charged pions, however, the contribution from the lowest Landau level to the pole masses, namely the term ${\cal F}^{11}_{\pi_\pm}|eB|$ in (\ref{pole}), leads to a rather strong magnetic field dependence of $m_{\pi_\pm}^{(0)}=m_{\pi_\pm}^{(z)}$. Taking into account the fact that $\pi_0$ is the lightest meson in the model and controls the interaction range between two quarks, we only consider its screening mass. In comparison with sigma and charged pions, the magnetic field dependence of the Goldstone mode is very weak, especially for its screening mass in the perpendicular direction.

\begin{figure}[H]
\centering
\includegraphics[height=0.50\textwidth]{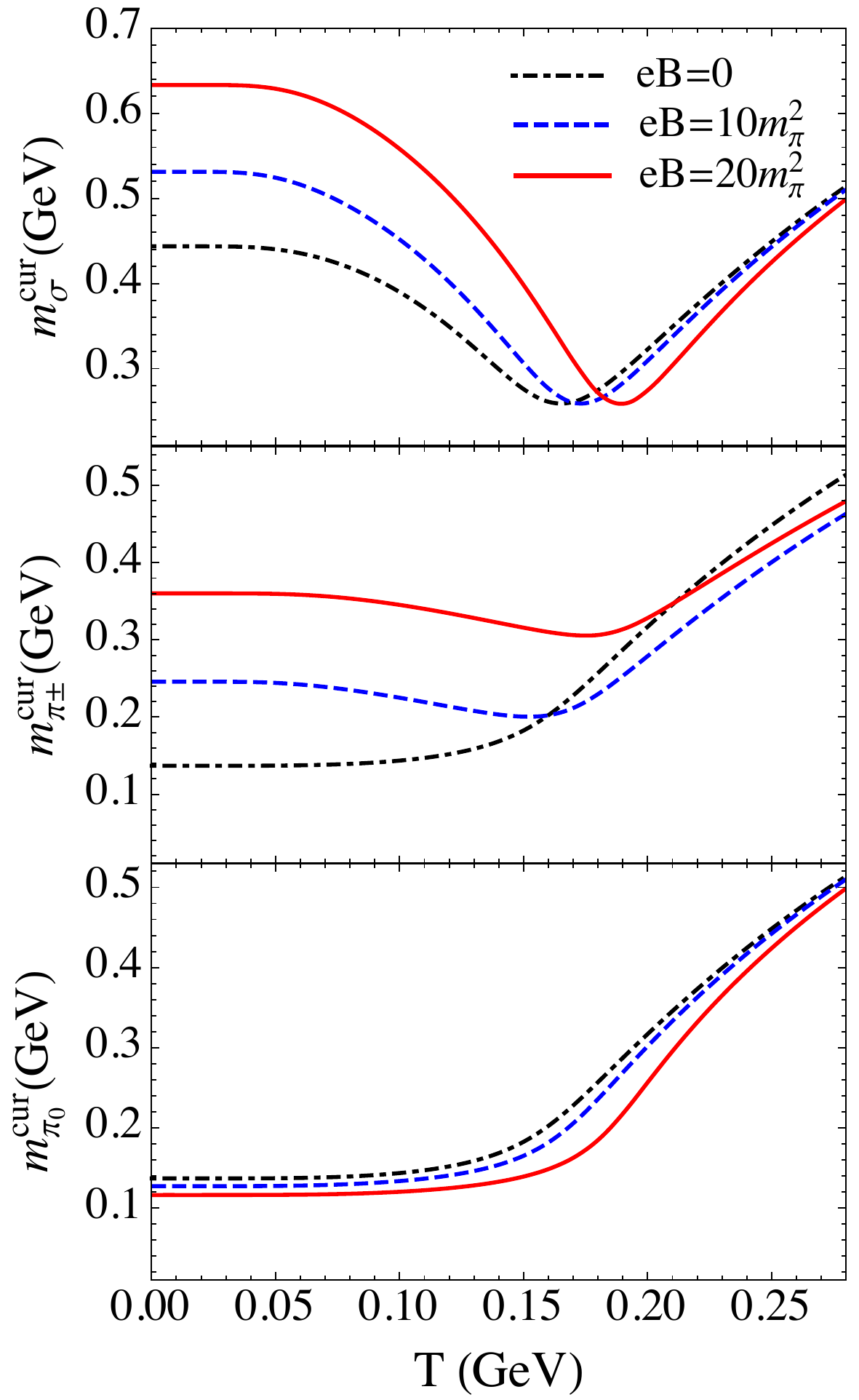}
\caption{The curvature masses at finite temperature and magnetic field for sigma (upper panel), charged pions (middle panel) and neutral pion (lower panel). }
\label{fig2}
\end{figure}

We now focus on the temperature dependence of the meson curvature masses which dominate the pole and screening masses. Fig.\ref{fig2} show $m_\sigma, m_{\pi_\pm}$ and $m_{\pi_0}$ as functions of temperature at fixed magnetic field $eB/m_\pi^2=0, 10$ and $20$. At low temperature, different from $\sigma$ and charged pions $\pi_\pm$ which obtain mass from the interaction with magnetic field, the Goldstone mode, namely the neutral pion $\pi_0$, loses mass in the interaction. This means that, the Goldstone mode controls the thermodynamics of the system and its importance increases with magnetic field. For the scalar meson $\sigma$ which is the Higgs mode corresponding to the chiral symmetry breaking, its mass continuously drops down at low temperature, reaches the minimum at the critical point of chiral phase transition, and then goes up in the phase with chiral symmetry restoration. At $B=0$ the critical temperature is $165$ MeV. At high temperature, the magnetic field dependence of any curvature mass becomes weak, they approach the same limit governed by the restored chiral symmetry.

\begin{figure}[H]
\centering
\includegraphics[height=0.35\textwidth]{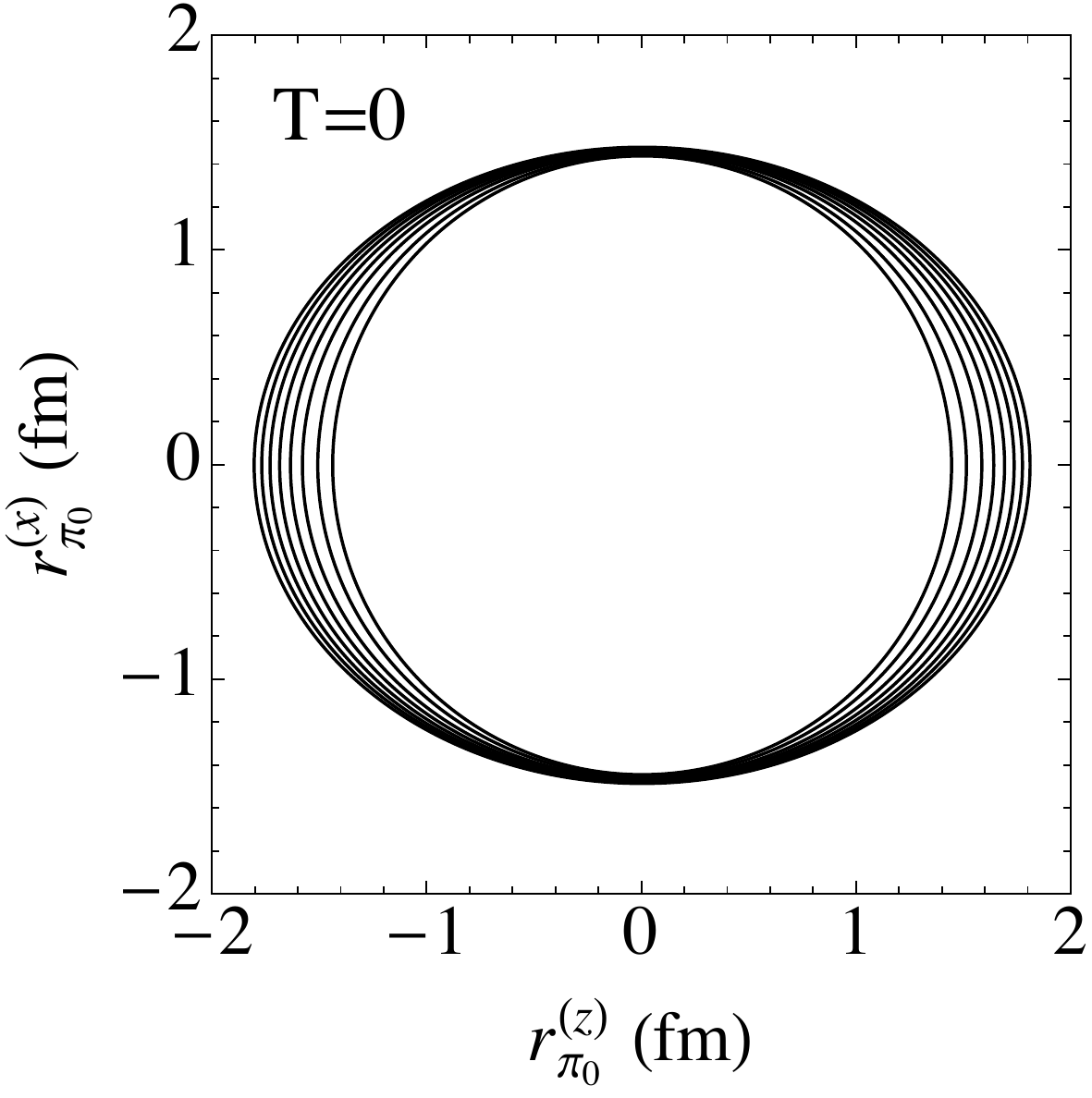}
\caption{The screening region of quark interaction in $z-x$ plane at zero temperature. $z$ and $x$ are the directions parallel and perpendicular to the magnetic field, and the fixed field strength corresponding to each ellipse increases from inside ($eB/m_\pi^2=0$) to outside ($20$). }
\label{fig3}
\end{figure}

The Goldstone mode controls not only the thermodynamics of the system but also the quark interaction propagated by mesons. As the lightest meson, $\pi_0$ governs the interaction range. The screening at finite temperature suppresses the long range part and remains only the short range part, the critical length is called screening radius determined by the screening mass, $r_{\pi_0}=1/m_{\pi_0}$. When the symmetry in coordinate space is broken by the magnetic field, the screening radius in the directions parallel and perpendicular to the field become different, and the interaction region changes from a sphere to a ellipsoid. From $m_{\pi_0}^{(x)} > m_{\pi_0}^{(z)}$ shown in the lower panel of Fig.\ref{fig1}, the short axis of the ellipsoid is in the perpendicular plane ($x-y$ plane) and the long axis is in the parallel ($z$) direction. Fig.\ref{fig3} shows the magnetic field dependence of the ellipse in the $z-x$ plane at zero temperature. Since $m_{\pi_0}^{(x)}$ is almost a constant in magnetic field, the short axis does not change. However, with increasing magnetic field the decreased screening mass $m_{\pi_0}^{(z)}$ leads to an increasing long axis. At finite temperature, the random thermal motion tends to recover the space symmetry, and the ellipsoid approaches back to the sphere.

\section{Conclusion}
\label{s5}
In this paper, we systematically study the neutral and charged meson properties starting with a quark model at finite temperature and magnetic field. Taking the bosonization method in a two-flavor NJL model, we derive the kinetic part of local meson Lagrangian from the non-local one-loop Lagrangian in the model by derivative expansion. The Schwinger phases in quark propagators caused by the magnetic potential are processed in a self-consistent way, leading to covariant derivatives acting on charged mesons. From the kinetic part of the effective meson Lagrangian, we extract the meson curvature masses and wave function renormalization constants which together control the meson pole and screening masses.

The magnetic field breaks down the isospin symmetry among the neutral and charged pions and the symmetry in coordinate space. As a result, the pion mass splits into two branches in magnetic field, and the wave function renormalizations in the directions parallel and perpendicular to the field are different. Due to the contribution from the lowest Landau level, the charged pions become much heavier in magnetic field and are sensitive to the field strength. As the only Goldstone mode corresponding to spontaneous chiral symmetry breaking, the neutral pion governs the thermodynamics of the system and propagates the long range quark interaction. From the magnetic field dependence of the neutral pion screening mass, the quark interaction region changes from a sphere in vacuum to a ellipsoid in magnetic field, and the asymmetry increases with the field strength.

\appendix {\bf Acknowledgement}: The work is supported by the NSFC and MOST grant Nos. 11335005, 11575093 and 2014CB845400.

\bibliographystyle{unsrt}

\end{document}